# Quantum interference of electromechanically stabilized emitters in nanophotonic devices

B. Machielse[1]*, S. Bogdanovic[2]*, S. Meesala[2]*, S. Gauthier[3], M. J. Burek[2], G. Joe[2], M. Chalupnik[1], Y. I. Sohn[2], J. Holzgrafe[2], R. E. Evans[1], C. Chia[2], H. Atikian[2], M. K. Bhaskar[1], D. D. Sukachev[1,4], L. Shao[2], S. Maity[2], M. D. Lukin[1], M. Loncar[2]†

[1] Department of Physics, Harvard University, Cambridge, MA 02138, USA
[2] John A. Paulson School of Engineering and Applied Sciences, Harvard University, Cambridge, MA 02138, USA
[3] Department of Physics and Astronomy, University of Waterloo, 200 University Avenue West
Waterloo, Ontario, Canada
[4] P. N. Lebedev Physical Institute of the RAS, Moscow 119991, Russia
* These authors contributed equally to this work
† Corresponding author. Email: loncar@seas.harvard.edu

**Photon-mediated coupling between distant matter qubits [1,2] may enable secure communication over long distances, the implementation of distributed quantum computing schemes, and the exploration of new regimes of many-body quantum dynamics [3,4]. Solid-state quantum emitters coupled to nanophotonic devices represent a promising approach towards these goals, as they combine strong light-matter interaction and high photon collection efficiencies [5-7]. However, nanostructured environments introduce mismatch and diffusion in optical transition frequencies of emitters, making reliable photon-mediated entanglement generation infeasible [7]. Here we address this long-standing challenge by employing silicon-vacancy (SiV) color centers embedded in electromechanically deflectable nanophotonic waveguides. This electromechanical strain control enables control and stabilization of optical resonance between two SiV centers on the hour timescale. Using this platform, we observe the signature of an entangled, superradiant state arising from quantum interference between two spatially separated emitters in a waveguide. This demonstration and the developed platform constitute a crucial step towards a scalable quantum network with solid state quantum emitters.**

Solid-state emitters with inversion symmetry [8] are promising for use in optical quantum networks due to their ability to be integrated into nanophotonic devices [9-11]. These emitters have suppressed static electric dipole moments, increasing the stability of their optical transition frequencies to fluctuations of electric fields that occur near device surfaces [8,12]. This fundamental property has been leveraged most notably with the negatively charged silicon-vacancy center in diamond (SiV) to achieve emitter-photon interaction with cooperativity greater than 20 in nanophotonic cavities [11]. The platform developed around this color center also provides other essential components for quantum networking, such as a long-lived quantum memory [10] and efficient photon collection [13].

Nevertheless, in terms of advancing towards multi-qubit networks, even inversion-symmetric emitters present challenges when incorporated into nanophotonic devices. They exhibit a significant inhomogeneous distribution of their optical transition frequencies as well as residual instability in the form of spectral diffusion [8,12]. Demonstrating full control of their spectral behavior is necessary to generate mutually indistinguishable photons, the key ingredient for long distance entanglement [14]. Because these emitters cannot be spectrally tuned using electric fields—the established tuning mechanism for solid-state emitters [7, 15]—previous experiments involving indistinguishable photon generation from such defects have relied on Raman [9, 16] and magnetic field [11] tuning. However, these techniques constrain either the spin or optical degree of freedom of the color center and are challenging to implement in the multi-qubit regime.

Inspired by experiments involving semiconductor quantum dots [17,18], initial experiments with diamond cantilevers have demonstrated the capacity to tune single SiV centers using strain [19, 20]. However, the suitability of this approach for photon mediated entanglement of multiple emitters remains an open question. In this work, we show that strain control can be used to manipulate the optical resonances of solid-state emitters and generate quantum interference between them. The demonstrated integration of strain control with nanophotonic devices represents a key step towards the realization of scalable quantum networks.

Our devices, presented in Figure 1, consist of triangular cross-section waveguides fabricated from single crystal diamond [21, 22]. Each of these waveguides is connected to a support structure on one end (Figure 1b). The other end is tapered to allow for collection of photons from the waveguide into an optical fiber with better than 85% efficiency [13] (Figure 1c). In order to embed SiV centers within diamond nanophotonic devices, we adapt a masked implantation technique previously used for bulk substrates [23, SI]. After the creation of SiV centers, gold electrodes are patterned onto the devices such that metallized parts of the waveguide act as one plate of a capacitor, with the other plate located on the diamond substrate (Figure 1b). Applying a voltage difference to these plates generates a force that deflects a portion of the waveguide, applying electrically controllable strain to the embedded SiV centers. This strain field perturbs the diamond lattice around the SiV which shifts the Coulomb energy of the orbitals and tunes the frequency of the optical transitions [8].

We measure the strain response of two SiV centers 30 μm apart within the same diamond device by resonantly exciting their optical transitions and collecting their phonon sideband emission at 4 K (Figure 2). The difference in the position of SiV centers in the device accounts for the difference in their response to the applied voltage [20], allowing us to overlap their optical transitions (Figure 2, red inset). We observe a tuning range of over 80 GHz, a factor of 3 larger than the total inhomogeneous distribution of the SiV optical transitions measured in these devices (Figure 2, blue shaded region). This is sufficient range to completely eliminate the effects of static strain variations on SiV optical transitions.

To determine the bandwidth of our electromechanical actuation scheme, we investigate an SiV color center's spectral response to AC mechanical driving of the nanophotonic structure (Figure 2, blue inset). We apply a DC bias combined with a variable-frequency RF signal to the gold electrodes and monitor the SiV optical transition. When the RF drive frequency matches one of the nanobeam's mechanical modes, we observe linewidth broadening of the SiV coupled to this mechanical mode due to the resonant amplification of the driving signal [25]. Using this technique, we observe and drive modes with mechanical frequencies of up to 100 MHz. Optimizing the device design could enable the electromechanical driving of vibrational modes with GHz frequencies which are resonant with SiV spin transitions. This could open new possibilities for engineering coherent spin-phonon interactions [19].

The high bandwidth of our electromechanical actuation scheme is sufficient to suppress spectral diffusion exhibited by the SiV center [12]. Monitoring an SiV center's optical transition over the course of five hours, we observe spectral diffusion that is an order of magnitude larger than its single-scan linewidth of around 300 MHz (Figure 3a). Using a pulsed feedback scheme that applies a voltage adjustment every 20 seconds we reduce the total summed linewidth by almost an order of magnitude on a timescale of several hours (Figure 3b). The efficacy of this feedback scheme is limited by the long duration of the laser scan used to capture the full extent of the spectral diffusion. This results in imperfect locking of the SiV optical transition to the target frequency (see SI).

The bandwidth of this spectral stabilization technique is improved by performing a faster 5 Hz sweep around a narrower region around the position of the SiV center optical transition. We monitor the noise spectrum of the SiV center emission with and without a lock-in feedback scheme applied to the device voltage (Figure 3c). The application of feedback results in an intensity noise reduction by approximately 8 dB, consistent with the stabilization achieved in Figure 3b. We attribute the frequency component dominated by the 1/f noise to slow strain fluctuations and second order susceptibility to electric field fluctuations in the SiV environment. Beyond this regime, our feedback scheme is limited by photon shot noise (Figure 3c). Further improving the SiV count rate through Purcell enhancement using a cavity could enable higher frequency feedback, as the existing scheme uses only a small share of the available actuation bandwidth.

With the optical transitions of the SiV centers tuned and stabilized, we proceed to generate probabilistic entanglement between two color centers. We begin by using strain control to set the optical transition frequencies between the lower branch of the ground state ($|c\rangle$) and the excited state ($|e\rangle$) of two emitters at the target detuning ($\Delta$) by applying the appropriate voltage (Figure 4a). The transitions between the upper branch of the ground state ($|u\rangle$) and the excited state ($|e\rangle$) of the two emitters are then continuously excited using two separate lasers (Figure 4a). Finally, the photons emitted into the diamond waveguide are collected through a tapered fiber interface and filtered using a high finesse Fabry-Perot filter (see SI) in order to separate the desired optical transition from the excitation laser [9].

We characterize our setup by measuring the second order correlation function between photons from a single excited SiV center (Figure 4c, left panel). We observe the suppression of photon coincidences with $g^{(2)}_{single}(0) = 0.13(4)$, a quantity raised above its ideal value of 0 by dark counts of the photon detectors. Next, we simultaneously excite both SiV centers and measure their photon correlation function. When two excited SiV centers are spectrally detuned (Δ ≠ 0), their emitted photons are distinguishable and the detection of the first photon projects the system into the statistical mixture of the |ce> and |ec> states (Figure 4b, blue path) [9]. In this case, the zero-time-delay photon correlation function reaches $g^{(2)}_{dist}(0) = 0.50(7)$ (Figure 4c, left panel).

When the two SiV optical transitions are tuned into resonance (Δ = 0), detection of the first emitted photon from a pair of excited SiV color centers projects the system into an entangled, bright state (Figure 4b, red path). This state is identified by the superradiant emission of a second photon at twice the rate expected from the distinguishable case [9, 26]. We confirm the generation of an entangled state by observing a superradiant peak in the photon correlation function (Figure 4c, right panel) with $g^{(2)}_{ind}(0) = 0.88(8)$. Experimental results are in good agreement with a simulated model of our system (Figure 4c, red curve) which uses only a single fitting parameter, the phonon-induced mixing rate between the |u> and |c> ground states [27, SI]. Using the height of the superradiant peak and the value of the $g^{(2)}_{single}$, we calculate a lower bound on the conditional bright state fidelity to be 0.8(1), indicating the observation of an entangled state [9].

In conclusion, we show a scalable approach to generating quantum interference between solid state emitters in nanophotonic devices using strain. This approach can be directly adapted to other quantum emitters such as quantum dots, as well as other inversion-symmetric color centers in diamond [28]. Furthermore, high frequency strain control inside a nanophotonic cavity can lead to a photon or phonon mediated gate between quantum memories [11, 29, 30]. Our platform fulfills the requirements for an ideal quantum network node and paves the way for realization of large-scale quantum networks.

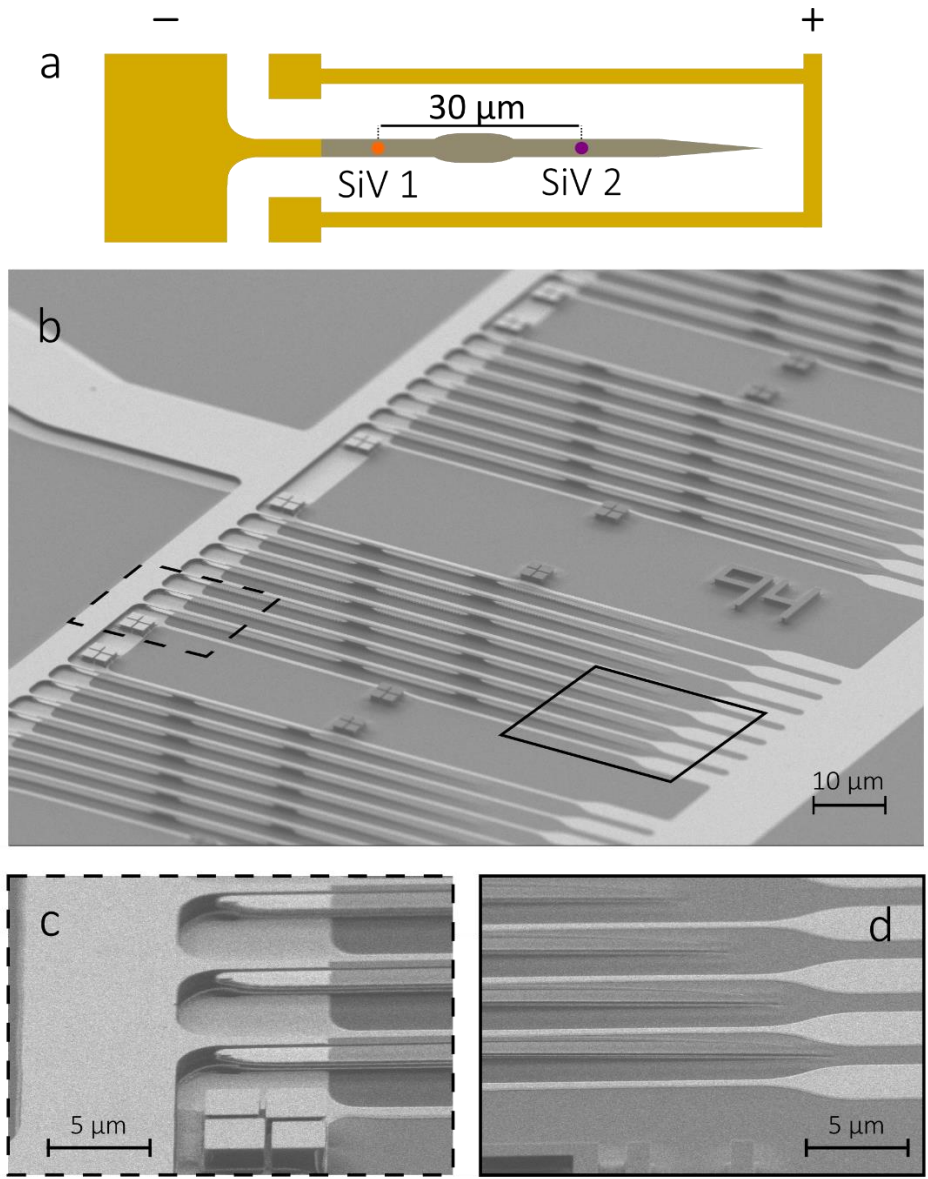

**Figure 1: Schematic of diamond nanophotonics device. a)** Diamond waveguides (gray) are implanted with color centers (purple and orange) at desired locations on the device. Electrodes (gold) are used to define a capacitor between plates located on the device (negative terminal) and below the device (positive terminal). Applying bias voltage between the plates causes the deflection of the doubly clamped cantilever. This tunes color centers between the plates and the first clamp (orange spot) without perturbing color centers beyond the clamp (purple spot). **b)** Scanning electron micrograph (SEM) of the photonic devices. **c)** Capacitor plates located on and below the devices. **d)** Diamond tapers used to extract photons from waveguides. This enables the extraction efficiency of more than 85% from the diamond waveguide into the fiber.

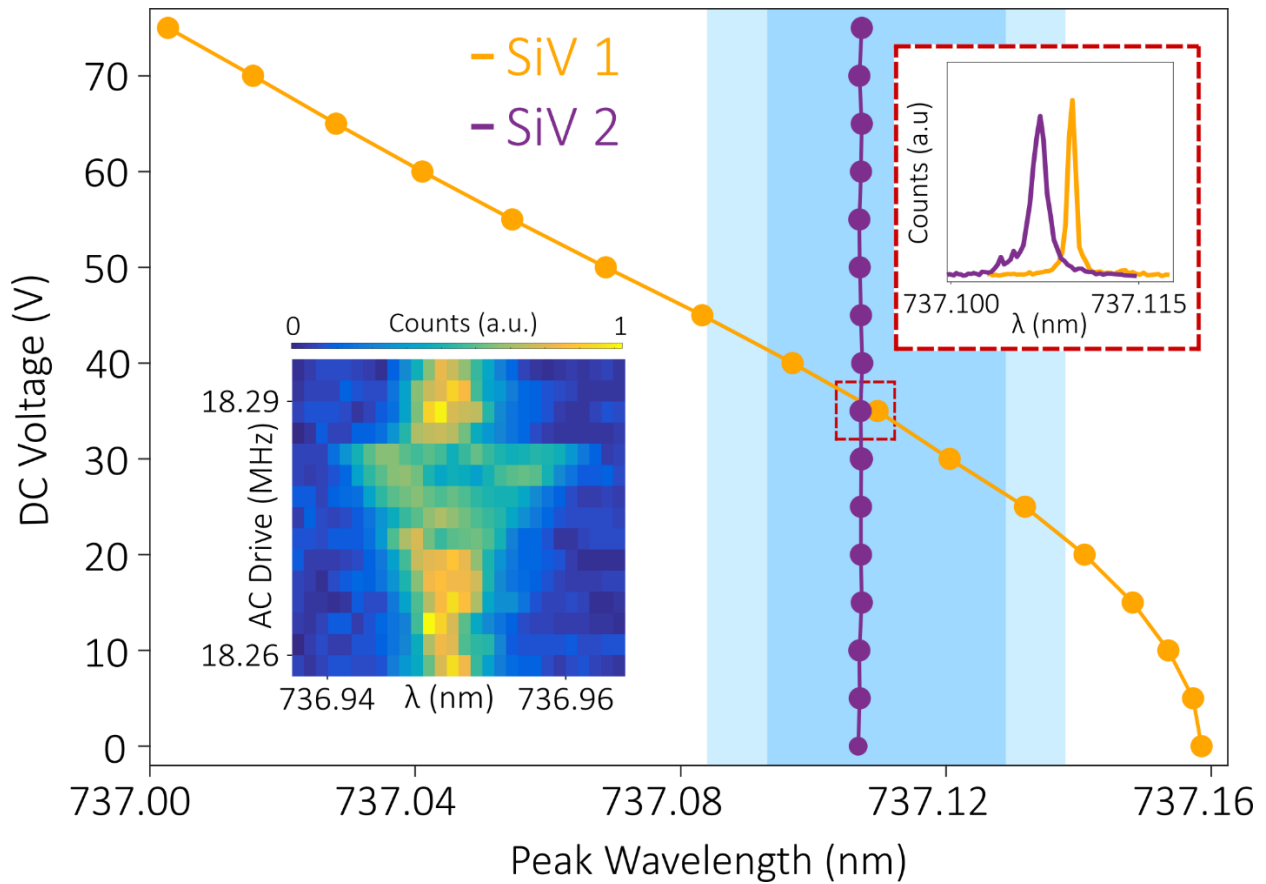

**Figure 2: Characterization of DC and AC voltage response of the devices**. Voltage bias is applied to the capacitor plates, resulting in strain fields that tune the optical transitions of SiV1 in the deflected portion of the device (orange) compared to the optical transitions of SiV2 in the stationary regions (purple). The tuning range of the SiV color center greatly exceeds the inhomogeneous distribution for 50% (75%) of SiV color centers observed in this experiment shown in dark (light) blue shading. The red inset shows the photoluminescence excitation spectra of two color centers at a voltage near the overlap. The blue inset shows the AC response of the cantilever system, measured by observing SiV optical transitions while modulating the AC driving frequency. Driving the cantilever resonance with its mechanical mode results in linewidth broadening of the color center optical transitions.

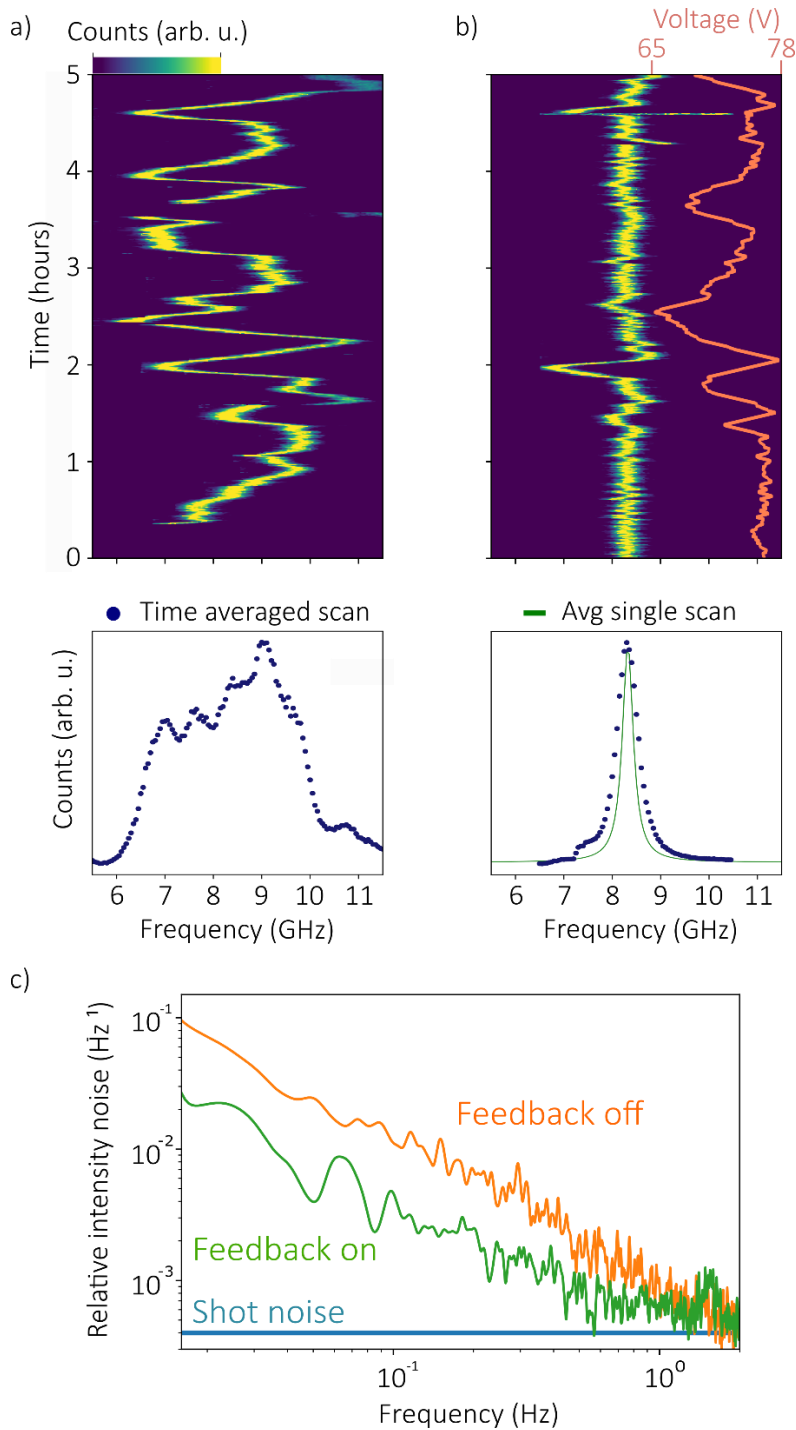

**Figure 3: Reduction of spectral diffusion using nanomechanical strain. a)** Top: Spectral diffusion of SiV center optical transition, measured over a 5-hour period. Bottom: Time-averaged spectrum over the same period. **b)** Top: Measurement of the SiV spectral diffusion with 20 second pulsed feedback over a 5-hour period. Bottom: Time-averaged spectra over a 5-hour period (blue dots) show the reduction of total linewidth down to 500 MHz. The average of single-scan SiV linewidths is 350 MHz (green line). **c)** PSD of measured SiV count fluctuations with (without) feedback is presented by the green (orange) line, showing reduction of noise by an average of 8 dB. Blue line indicates statistical noise limit set by finite photon emission and collection rate.

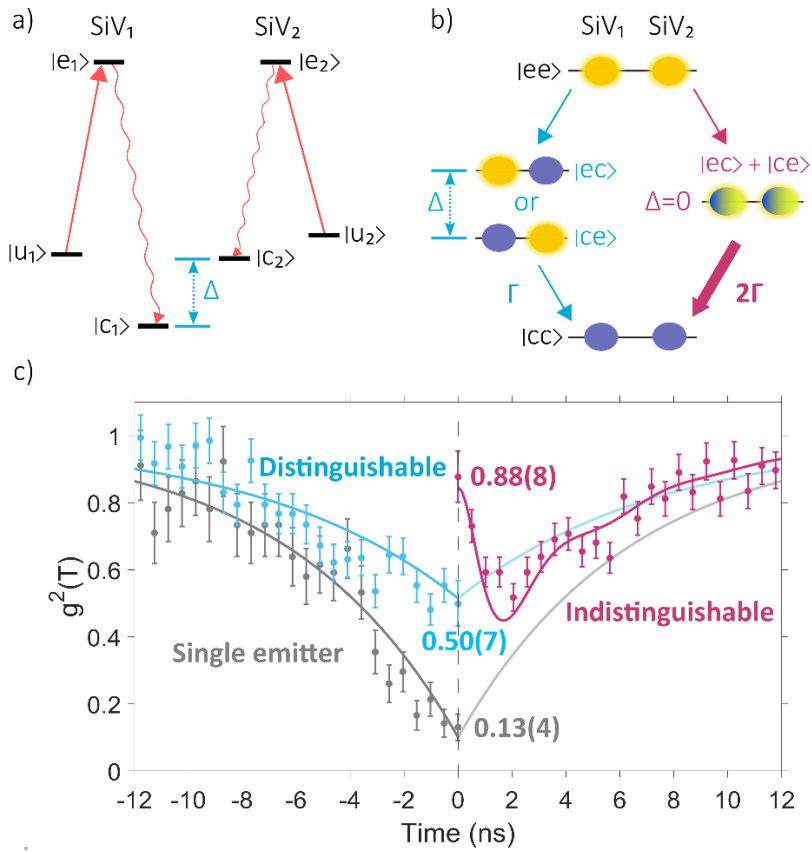

**Figure 4: Generation of the superradiant entangled state using strain control. a)** Level structure for two SiV centers. Strain tuning of one SiV color center changes the detuning (Δ) between their |e> to |c> transitions. Separate lasers are applied to excite the |u> to |e> transition of each emitter. **b)** A single photon emitted from the two excited emitters with distinguishable transitions (Δ ≠ 0) projects the system to a statistical mixture of |ec> and |ce> states (blue decay path). When the emitters are indistinguishable (Δ = 0), the emission of one photon projects the system into a superradiant bright state (purple decay path) that decays at a rate two times faster than that of the statistical mixture. **c)** The second order photon correlation function is measured for a single emitter (left panel, gray data points) and for two spectrally distinguishable emitters (left panel, blue data points). Measured $g^{(2)}(0)$ is 0.13(4) and 0.50(7) for the single emitter and distinguishable cases, respectively. Exponential curves fit to the data are plotted and mirrored onto the right half of plot (gray and blue lines). For two emitters tuned into resonance, we observe the generation of a superradiant entangled state, signified by the peak in the photon correlation (right panel, purple data points). Data is overlaid with a simulated model with a single fit parameter (see SI). For indistinguishable emitters, we measure $g^{(2)}(0)$=0.88(8), limited primarily by detector jitter.